\documentclass[conference]{IEEEtran}

\usepackage{amsmath}
\usepackage[tight]{subfigure}
\usepackage[footnotesize]{caption} 
\usepackage{cite}
\usepackage{eurosym}
\usepackage{amsfonts}
\usepackage{graphics}
\usepackage{epsfig}
\usepackage{psfrag}
\usepackage{pstricks}
\usepackage{array}
\usepackage{shortvrb}
\usepackage{epsf}
\usepackage{graphicx}
\usepackage{rotating}
\usepackage{float}
\usepackage{color}
\usepackage{cite}
\usepackage{soul}
\usepackage{multirow}
\usepackage{acronym}
\usepackage{url}
\usepackage{breakurl}
\usepackage[breaklinks, hidelinks]{hyperref}

\acrodef{ess}[\textsc{ess}]{Energy Storage System}
\acrodef{dae}[\textsc{dae}]{Differential Algebraic Equation}
\acrodef{vsc}[\textsc{vsc}]{Voltage Sourced Converter}
\acrodef{coi}[\textsc{coi}]{Centre of Inertia}
\acrodef{fdf}[\textsc{fdf}]{Frequency Divider Formula}
\acrodef{wecs}[\textsc{wecs}]{Wind Energy Conversion System}
\acrodef{spvg}[\textsc{spvg}]{Solar Photo-Voltaic Generation}
\acrodef{tcl}[\textsc{tcl}]{Thermostatically Controlled Load}
\acrodef{hvdc}[\textsc{hvdc}]{High-Voltage Direct Current}
\acrodef{pll}[\textsc{pll}]{Phase-Locked Loop}
\acrodef{pmu}[\textsc{pmu}]{Phasor Measurement Unit}
\acrodef{rtds}[\textsc{rtds}]{Real-Time Digital Simulator}
\acrodef{emt}[\textsc{emt}]{Electromagnetic Transients}
\acrodef{tg}[\textsc{tg}]{Turbine Governor}
\acrodef{avr}[\textsc{avr}]{Automatic Voltage Regulator}
\acrodef{agc}[\textsc{agc}]{Automatic Generation Control}
\acrodef{gps}[\textsc{gps}]{Global Positioning Satellite}

\DeclareGraphicsExtensions{.eps,.png,.pdf}

\def \R {{\rm I\kern -2.2pt R\hskip 1pt}}

\newcommand{\PreserveBackslash}[1]{\let\temp=\\#1\let\\=\temp}

\IEEEoverridecommandlockouts
\begin{document}

\title{Emerging Challenges of Integrating Solar PV in the Ireland and Northern Ireland Power Systems}

\author{ \IEEEauthorblockN{Taulant K\"{e}r\c{c}i,\IEEEauthorrefmark{1}
    \IEEEmembership{IEEE~Member}, Manuel Hurtado,\IEEEauthorrefmark{1} \IEEEmembership{IEEE~Member}, Simon Tweed,\IEEEauthorrefmark{1} Marta Val Escudero,\IEEEauthorrefmark{1} \\Eoin Kennedy,\IEEEauthorrefmark{1}
    and
    Federico~Milano,\IEEEauthorrefmark{2}~\IEEEmembership{IEEE~Fellow}}\vspace*{0.3cm}
  \IEEEauthorblockA{
    \begin{tabular}{cc}
      \begin{tabular}{@{}c@{}}
        \IEEEauthorrefmark{1}
        Transmission System Operator \\
        Innovation \& Planning Office, EirGrid, plc\\
        Ireland
      \end{tabular} &
      \hspace{0.3cm}
      \begin{tabular}{@{}c@{}}
        \IEEEauthorrefmark{2}
        School of Electrical and Electronic Engineering \\ University College Dublin \\Ireland
      \end{tabular} 
    \end{tabular}
  }
  %
  %
}

\IEEEoverridecommandlockouts

\maketitle
\IEEEpubidadjcol

\begin{abstract}
  This paper discusses emerging operational challenges associated with the integration of solar photovoltaic (PV) in the All-Island power system (AIPS) of Ireland and Northern Ireland.  These include the impact of solar PV on: (i) dispatch down levels; (ii) long-term frequency deviations; (iii) voltage magnitude variations; and (iv) operational demand variations.  A case study based on actual data from the AIPS is used to analyze the above challenges.  It is shown that despite its (still) relatively low penetration compared to wind power penetration, solar PV is challenging the real-time operation of the AIPS, e.g., maintaining frequency within operational limits.  EirGrid and SONI, the transmission system operators (TSOs) of the AIPS, are working toward addressing all the above challenges. 
\end{abstract}

\begin{IEEEkeywords}
  Solar PV, integration, challenges, variability, frequency variations, operational demand.
\end{IEEEkeywords}

\section{Introduction}
\label{sec:intro}

\subsection{Motivation}
\label{Motivation}

EirGrid and SONI, the transmission system operators (TSOs) of Ireland (IE) and Northern Ireland (NI), respectively, have successfully integrated high levels of variable non-synchronous renewable energy sources (RESs), notably wind power (approximately 6 GW installed as of 2022).  Regarding instantaneous system non-synchronous penetration (SNSP), the TSOs accommodate up to 75\% at any point in time and plan to raise this limit to 95\% by 2030 \cite{eirgrid}.  Despite these unprecedented levels of wind power, the governments of IE and NI have set ambitious targets to integrate high shares of solar photovoltaic (PV) as well.  For example, the latest IE climate action plan (CAP) foresees up to 5 GW and 8 GW solar PV capacity by 2025 and 2030, respectively \cite{gov}.  The distribution system operator (DSO) in IE, in fact, expects to have over 1 GW of distributed-connected solar in IE by the end of 2023, making it the fastest-growing renewable industry \cite{solar}.  These solar PV levels have started introducing challenges such as maintaining frequency, voltage and operational demand within limits and managing its dispatch down levels.  This paper aims to analyze these challenges using actual data and provide insights into how those can be addressed.  

\begin{figure}[t!]
  \begin{center}
    \resizebox{0.9\linewidth}{!}{\includegraphics{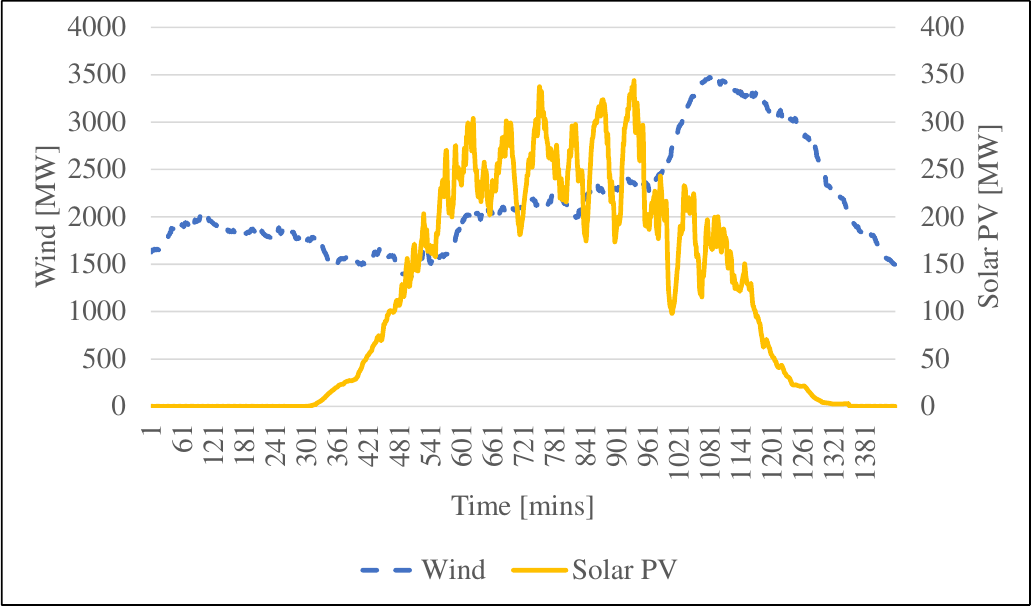}}
    \caption{Typical daily solar PV and wind power profiles in the AIPS.}
    \label{fig:variability}
  \end{center}
  \vspace*{-0.3cm}
\end{figure}

\subsection{Literature Review}
\label{sec:literature}

Real-world experiences show that due to cloud passing, solar PV output can drop as high as 63\% of the rated capacity/minute \cite{6763050}.  This variability is significantly higher compared to, for example, wind power \cite{ZHANG2019809}.  To illustrate these differences, we plot in Fig.~\ref{fig:variability}  typical solar PV and wind power profiles in the All-Island power system (AIPS).  It can be seen that solar PV generation is much more volatile throughout the day.  As it will be shown later in the paper, such volatility, despite still low in magnitude, introduces challenges in terms of frequency and voltage management, among others \cite{9773105}.  

While the literature is rich in studying the impact of solar PV on the transient frequency response of power systems \cite{8273793}, that is not the case for its impact on long-term frequency deviations.  An exception is reference \cite{yuan2020multi}, where the authors demonstrate that a value of 3\% of the regulation reserve might not be enough to keep frequency within operational limits during a cloudy day.  Similarly, reference \cite{marta} uses a 2040 transmission system model of IE and shows that while frequency remains within normal operating range (49.80 Hz to 50.2 Hz), its standard deviation increases to 105\% when moving from 60\% and 80\% RESs penetration levels.

Another challenge associated with high shares of PV penetration is managing its curtailment.  For example, significant levels of PV curtailment (greater than 1\% of potential output) have been recorded in Chile, China, Germany, and specific markets in the United States \cite{OSHAUGHNESSY20201068}.  The issue of distributed PV curtailment in low voltage networks is discussed in \cite{YILDIZ2023113696} using real-world data from Australia.  Indeed, the curtailment may be the last action for TSOs to manage the minimum operational demand problem (i.e., the ``duck curve'') \cite{HOU2019205}.  On the other hand, the authors in \cite{LIU2023109554} stress the need to convert solar PV to a dispatchable source to address its intermittency nature.  Other solutions to reduce solar PV curtailment proposed in the literature are demand-side management and economic dispatch \cite{SAMBASIVAM2023113334}. 

\subsection{Contributions}
\label{Contributions}

To the author's knowledge, this is the first research paper to study the technical challenges of solar PV integration in the AIPS.  Most importantly, it does so based on a real-world power system and using actual data.  In addition, there is a lack of studies in the literature that focus on the impact of solar PV penetration on long-term frequency deviations.  In this context, this paper brings the following specific contributions:
\begin{itemize}
\item An analysis of four critical operational challenges related to the integration of solar PV in a real-world, large-scale renewable-dominated power system namely, the AIPS.
\item Demonstrate based on actual data that solar PV is challenging the operation of the AIPS despite its still relative low penetration when compared to wind power, e.g., keeping frequency and voltage within limits.
\end{itemize}

\subsection{Paper Organization}

The rest of the paper is structured as follows.  Section \ref{sec:background} provides a short background on the AIPS.  Section \ref{sec:case} discusses the four operational challenges based on actual measurement data.  Section \ref{sec:conclu} draws the main conclusions of the paper. 

\section{Background on the All-Island Power System}
\label{sec:background}

Both IE and NI have set ambitious renewable energy targets for 2030. The IE CAP set a target of 80\% of electricity met by renewable energy by 2030, while NI has a target of at least 70\% by 2030 \cite{gov}.  In IE and NI, renewable energy is predominantly sourced from wind, although solar energy has grown in size and significance in recent years.  For instance, in IE, there is around 309 MW and 700 MW of transmission-connected and DSO-connected solar generation, respectively \cite{statistics, solar}.  This includes 371 MW of utility-scale solar PV and almost 60,000 micro-generation customers \cite{solar}.  While in NI, there is approximately 182 MW of large-scale solar PV as of the end of 2022 \cite{statistics}, and a couple of hundreds MW of distributed PV (roof-top PV) installed \cite{statistics1}.  

Although current IE and NI solar PV capacity is still significantly lower than wind power, it is still introducing challenges such as: (i) dispatch down levels; (ii) long-term frequency deviations; (iii) voltage magnitude variations; and (iv) operational demand variations.  In particular, if certain operational constraints are binding, such as SNSP, then the TSOs instruct solar PV units to dispatch down their generation.  Specifically, the AIPS has in place four operational constraints/limits that impact solar PV dispatch down levels (see evolution of constraints in Tab.~\ref{tab:constraint}) namely \cite{eirgrid}: (i) SNSP; (ii) a minimum number of conventional units online (MUON); (iii) rate
of change of frequency (RoCoF) limit; and (iv) minimum inertia floor.  The interested reader is referred to \cite{10253224} for further information on each constraint above.

\begin{table}[h!]
  \centering
  \caption{Evolution of operational policy constraints in the AIPS \cite{eirgrid}.} 
  \label{tab:constraint}
  \begin{tabular}{cccccc}
    \hline
    Year & SNSP & RoCoF & Inertia & MUON  
     \\
    \hline
    2023  & 75\% & 1 Hz/s & 23 GWs & 7 \\
    2030 & 95\% & 1 Hz/s & 20 GWs & 3 \\
    \hline
  \end{tabular}
\end{table}

\section{Case Study}
\label{sec:case}

This section discusses the four operational challenges based on actual data obtained from the TSO SCADA.  Unless stated otherwise, frequency data sampling rate is seconds whereas solar PV generation sampling rate is minutes.  

\subsection{Dispatch Down}
\label{sec:curtail}

\begin{figure}[h!]
  \begin{center}
    \resizebox{0.9\linewidth}{!}{\includegraphics{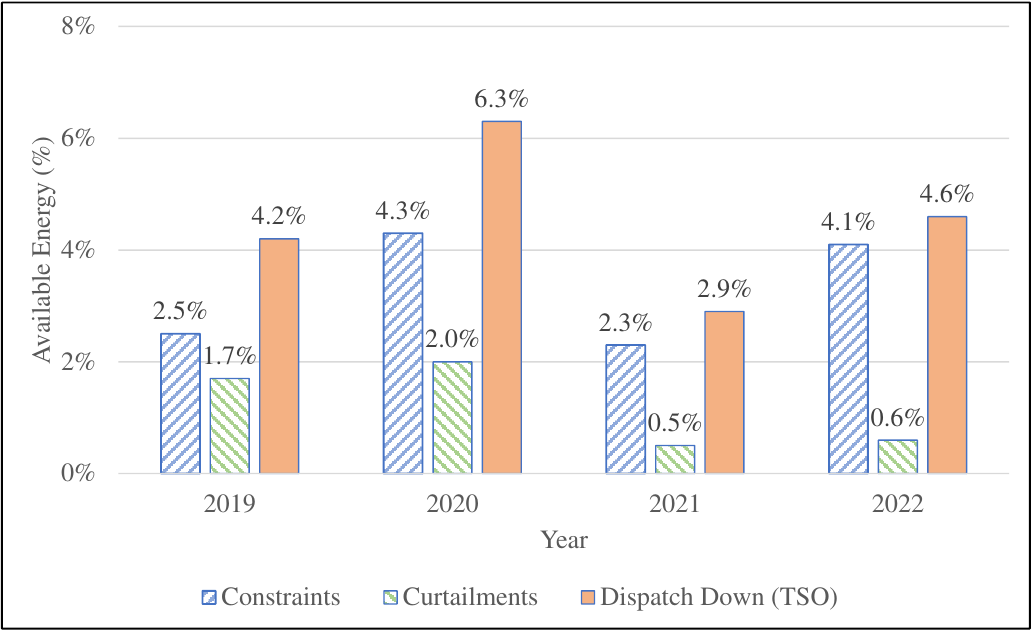}}
    \caption{Solar PV dispatch down over the recent years in NI.}
    \label{fig:dd}
  \end{center}
  \vspace*{-0.3cm}
\end{figure}

This section analyses solar PV dispatch down levels over the recent years in AIPS.  The focus is on NI as large-scale PV installations in IE started just recently (early 2023).  In the AIPS, the term dispatch down refers to instructions issued by TSOs to RESs (wind and solar) to reduce their generation output for localised network reasons (constraints) and system-wide reasons (curtailment).  Note that currently wind and solar PV are treated as priority dispatch.  Figure~\ref{fig:dd} shows dispatch down levels (in \%) over the last years.  It can be seen that there have been relevant dispatch down of solar PV.  These levels are lower than the dispatch down of wind \cite{10253224}.  However, note that apart from the lower PV penetration level, another main reason for these differences is that wind is generally being curtailed during night hours (low demand) while solar availability is during daytime hours (10:00 to 16:00) when the system conditions are such that allow to integrate more power into the grid, i.e., higher demand.

\begin{table}[h!]
  \centering
  \caption{Reason codes for solar PV curtailment in the AIPS.} 
  \label{tab:reason}
  \begin{tabular}{cccccc}
    \hline
    Year & SNSP & RoCoF/Inertia & High Freq/MUON  
     \\
    \hline
    2019  & 20\% & 0\% & 80\% \\
    2020 & 45\% & 0\% & 55\% \\
    2021  & 39\% & 0\%  & 61\%\\ 
    2022  & 5\% & 0\% & 95\%  \\ 
    \hline
  \end{tabular}
\end{table}

Figure~\ref{fig:dd} also depicts the split between constraints and  curtailments.  Currently dispatch down of solar PV mainly happens due to local network reasons (constraints).  As part of Shaping Our Electricity Future (SOEF) Roadmap, the TSOs are currently implementing a flexible network strategy, e.g., dynamic line rating, that together with the planned network investments, among others, will help reduce the network constraints and, thus, make more room for RESs \cite{soef}.  Table~\ref{tab:reason}, on the other hand, shows the sub categories of curtailment for the same period.  The dominant reason for PV curtailment is the MUON limit and the high-frequency challenge.  Indeed, this challenge is expected to grow in the future with higher shares of PV installation in the AIPS (see next section).  

\subsection{Impact on Frequency Quality}
\label{sec:frequency}

Until now, the TSOs have mainly been concerned with the impact of wind power penetration on the AIPS.  For example, more than a decade ago, the TSOs carried our various wind integration studies and identified different potential issues with high values of SNSP limit (e.g., 75\%).  Two of the main issues identified were RoCoF ($\pm 0.5$ Hz/s limit back at the time) and large wind ramps.  To address these issues, the TSOs introduced relevant system services such as fast frequency response and ramping margin products, among others \cite{control}.   

\begin{figure}[t!]
  \begin{center}
    \resizebox{0.9\linewidth}{!}{\includegraphics{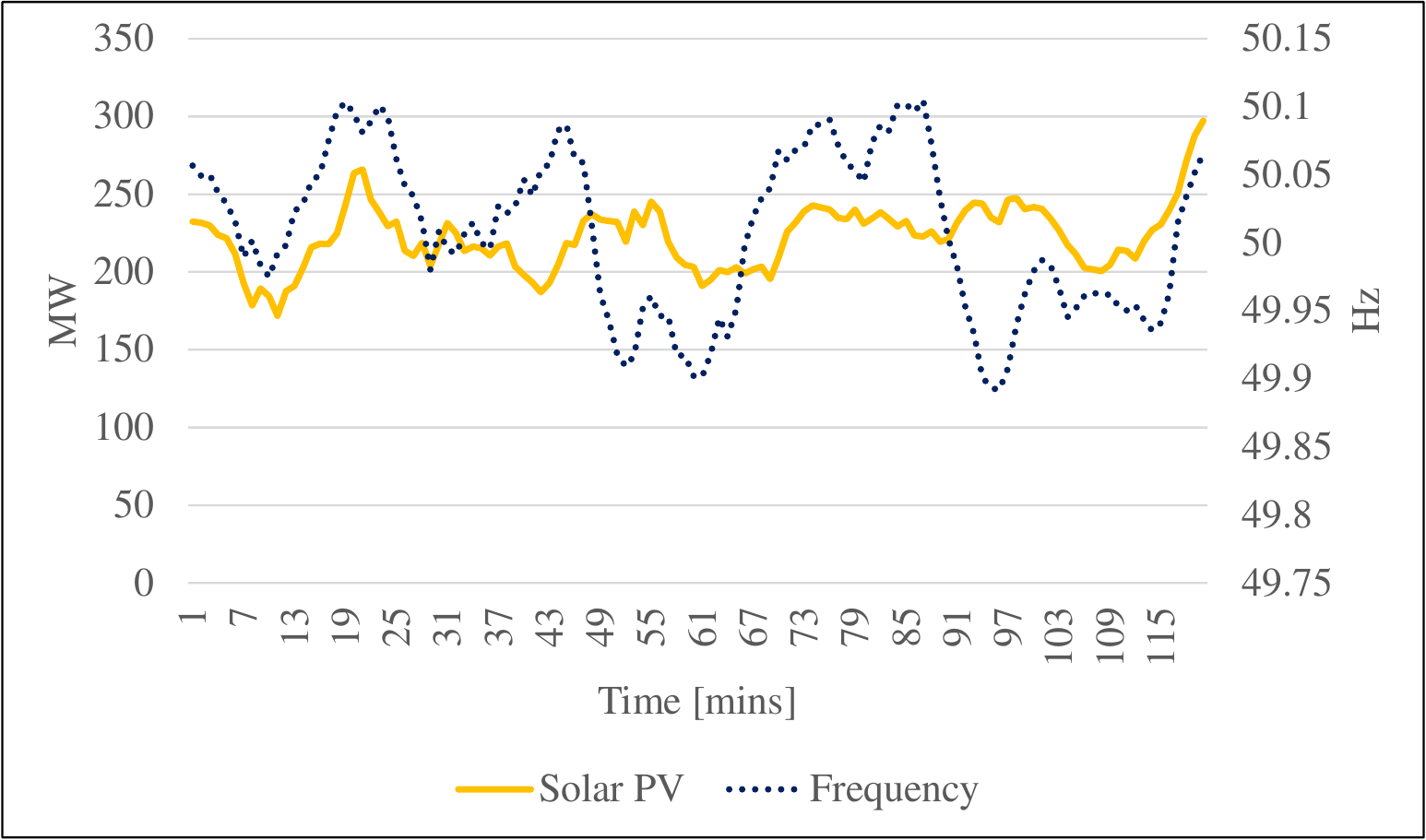}}
    \caption{Frequency and solar PV traces for two relevant hours in the AIPS.}
    \label{fig:freq}
  \end{center}
  \vspace*{-0.3cm}
\end{figure}

An emerging issue is higher frequency deviations due to increased PV penetration.  PV power profiles like that in Fig.~\ref{fig:variability} create additional challenges for control room operators.  In particular, it is becoming increasingly difficult to manage frequency within operational limits (e.g., $\pm 200$ mHz) when PV output power changes quickly due to cloud passes and/or changes in radiation level.  We illustrate this issue in Fig.~\ref{fig:freq}, where solar PV variations and frequency are plotted against each other for two hours of a relevant day.  It is interesting to see that in the first and last 30 minutes, there is an almost perfect linear relationship between solar and frequency variations.  However, when the frequency is about to drift outside the $\pm 100$ mHz range, then the control room operators implement manual operations (e.g., conventional generation redispatch) to bring back system frequency within limits.  
\begin{table}[t!]
  \centering
  \caption{Solar and wind share (\% of demand), PV dispatch down, and the violation minutes ($\pm$ 100 mHz criteria $\ge$ 98\% of time) for the period April - August for years 2022 and 2023. } 
  \label{tab:kpi}
  \begin{tabular}{cccccc}
    \hline
    Month & April & May & June & July & August   
     \\
    \hline
    $\rm P_{Solar\%}$ 2022  & 0.5 & 0.6 & 0.6 & 0.5 & 0.6 \\
    $\rm P_{Wind\%}$ 2022  & 32 & 34  & 30 & 21 & 20\\ 
    Dispatch Down 2022  & 6.4\% & 3.4\% & 6.3\% & 4.1\% & 4.2\% \\
    Violation Minutes 2022  & 411 & 433 & 333 & 230 & 316 \\
    \hline
    $\rm P_{Solar\%}$ 2023 & 1.4 & 2.6 & 2.8 & 2.4 & 2.2 \\
   $\rm P_{Wind\%}$ 2023  & 35 & 21 & 21 & 33 & 33  \\ 
   Dispatch Down 2023  & 2.9\% & 3.9\% & 10.6\% & 11.3\% & 11.4\% \\
    Violation Minutes 2023  & 486 & 348 & 494 & 626 & 624 \\
    \hline
  \end{tabular}
\end{table}

Table \ref{tab:kpi} compares the violation minutes for April to August 2022 and 2023.  During this period, the solar PV share in the system rose from 0.5\% to 2.5\% on average but, of course, it is still much lower than wind (e.g., around 20-30\% on average).  However, due to increased solar PV variations, frequency is spending more time outside the $\pm 100$ mHz range.  It is counter-intuitive that low solar penetration impacts more on frequency deviations than high wind generation.  However, what matters is not the average power, but the variance of such a power.  The low level of solar PV penetration, in fact, comes with a high volatility as PV power plants in the AIPS are currently located in few specific areas of the network.  Thus, the averaging effect that reduces the variance of wind-generated power cannot take place for solar generation.  This also suggests a way to smooth and reduce solar PV system output variability, that is, through geographical aggregation, namely, an even geographical installation of solar power plants in the whole network \cite{ALDEMAN2023100052}.   The increase in solar PV penetration not accompanied by geographical aggregation is one of the main factors leading to almost doubling the violation minutes in 2023.  A crucial frequency service that can address this problem is to turn on active power control for wind and solar plants \cite{10253411}.
Another important operational policy change that has impacted the increased number of violation minutes is the MOUN limit change from 8 to 7 (end of May 2023).  Operating with one conventional unit less in the system means there is less regulation reserve provided (by means of $\pm 15$ mHz deadband) on average and, in turn, a deterioration of frequency quality \cite{10253411}.  


\subsubsection{Pearson’s correlation coefficient}
\label{sec:pearson}

Similar to \cite{8783475}, we use the Pearson’s correlation coefficient to provide a more quantitative analysis of the impact of solar PV penetration on long-term frequency deviations in the AIPS.  The Pearson’s correlation coefficient is a statistical tool that measures the (linear) correlation between two variables.  Its formulation is:
\begin{align}
  \label{solar} 
  \rm r = \frac{\rm \sum_{i}^{N} (X_{i}-\overline X) (Y_{i}-\overline Y)}{\rm \sqrt{\sum_{i}^{N} (X_{i}-\overline X)^2 \sum_{i}^{N} (Y_{i}-\overline Y)^2}} \, ,
\end{align}
where $\rm N$ is the number of observations; $\rm X_i$ and $\rm Y_i$ are the values of the two time series, with length $\rm N$, whose correlation is to be calculated; $\rm \overline X$ and $\rm \overline Y$ are the mean values of the time
series $\rm X_i$ and $\rm Y_i$ , respectively;  $\rm r$ can take values from $-1$ to $1$.  If $\rm r = \pm 1$ exactly, then the relationship between, for example, the solar PV penetration (\% of demand) $\rm P_{Solar\%}$ and, say, the standard deviation of the frequency, $\sigma_f$, can be described by means of a linear equation.  On the other hand, $\rm r = 0$ indicates that there is no linear relationship between $\rm P_{Solar\%}$ and $\sigma_f$.  Further, if $\rm r > 0$, means that if $\rm P_{Solar\%}$ increases then also $\sigma_f$ increases and vice-versa if $\rm r < 0$.

The Pearson’s correlation coefficients are calculated taking $\rm X = \rm P_{Solar\%}$, i.e., the instantaneous value of solar energy as percentage share of system demand:
\begin{align}
  \label{solar} \rm P_{Solar\%} = \frac{\rm Averaged \; Solar}{\rm Averaged \; Demand} \, \cdot 100,
\end{align}
and ${\rm Y} = \sigma_f$, i.e., the standard deviation of the system frequency over the same period for which $\rm P_{Solar\%}$ is calculated.  For comparison, we calculate the Pearson’s correlation coefficients for wind and SNSP as well ($\rm X = \rm P_{Wind\%}$ and $\rm X = \rm SNSP$, and Y = $\sigma_f$) with wind energy share equation as follows:
\begin{align}
  \label{wind} \rm P_{Wind\%} = \frac{\rm Averaged \; Wind}{\rm Averaged \; Demand} \, \cdot 100.
\end{align}
Since large-scale PV installations in IE went live mostly around April-May 2023, we perform the analysis for months April-August 2023 \cite{statistics}.  Next, similar to \cite{8783475}, the focus is on the day hours, that is, the period from 10:00 to 16:00, in order to minimize the effect of load ramping.

\begin{table}[t!]
  \centering
  \caption{Pearson’s coefficients for Psolar\%, Pwind\%, SNSP\% and $\sigma_f$, respectively, for the AIPS in the period April - August 2023 and using 15-minute resolution.} 
  \label{tab:15minute}
  \begin{tabular}{cccccc}
    \hline
    Month & $\rm r_{Solar}$ & $\rm r_{Wind}$ & $\rm r_{SNSP}$   
     \\
    \hline
    April  & $\phantom{-}0.0573$ & $\phantom{-}0.1006$ & $\phantom{-}0.1125$ \\
    May & $-0.1483$ & $\phantom{-}0.3323$ & $\phantom{-}0.2857$ \\
    June  & $- 0.1766$ & $\phantom{-}0.3257$  & $\phantom{-}0.3074$ \\ 
    July  & $\phantom{-}0.1472$ & $- 0.0852$ & $- 0.0631$  \\ 
    August  & $\phantom{-}0.1056$ & $\phantom{-}0.1825$ & $\phantom{-}0.1935$ \\ 
    \hline
  \end{tabular}
\end{table}

\begin{table}[t!]
  \centering
  \caption{Pearson’s coefficients for Psolar\%, Pwind\%, SNSP\%, $\rm \sigma_{Solar}$, $\rm \sigma_{Wind}$ and $\sigma_f$, respectively, for the AIPS in the period April - August 2023 and using 5-minute resolution.} 
  \label{tab:5minute}
  \begin{tabular}{cccccc}
    \hline
    Month & $\rm r_{Solar}$ & $\rm r_{Wind}$ & $\rm r_{SNSP}$ &  $\rm r_{\sigma_{Solar}}$ & $\rm r_{\sigma_{Wind}}$  
     \\
    \hline
    April  & $\phantom{-}0.0436$ & $\phantom{-}0.0974$ & $\phantom{-}0.1008$ & $0.1460$  & $0.3975$ \\
    May & $-0.1104$ & $\phantom{-}0.3240$ & $\phantom{-}0.3058$ & $0.2170$ & $0.2650$ \\
    June  & $-0.1152$ & $\phantom{-}0.3212$  & $\phantom{-}0.3324$ & $0.2088$ & $0.2894$ \\ 
    July  & $\phantom{-}0.1225$ & $-0.0819$ & $-0.0630$ & $0.2385$ & $0.2799$  \\ 
    August  & $\phantom{-}0.0582$ & $\phantom{-}0.1571$ & $\phantom{-}0.1551$ & $0.2444$ & $0.4097$ \\ 
    \hline
  \end{tabular}
\end{table}
\begin{table}[t!]
  \centering
  \caption{Pearson’s coefficients for Psolar\%, Pwind\%, SNSP\% and $\sigma_f$, respectively,
for the AIPS in
the period April - August 2023 and using 1-minute resolution.} 
  \label{tab:1minute}
  \begin{tabular}{cccccc}
    \hline
    Month & $\rm r_{Solar}$ & $\rm r_{Wind}$ & $\rm r_{SNSP}$   
     \\
    \hline
    April  & $-0.0274$ & $\phantom{-}0.0739$ & $\phantom{-}0.0505$ \\
    May & $-0.1343$ & $\phantom{-}0.2640$ & $\phantom{-}0.2353$ \\
    June  & $-0.0569$ & $\phantom{-}0.3073$  & $\phantom{-}0.3537$ \\ 
    July  & $-0.0079$ & $-0.0660$ & $-0.0607$  \\ 
    August  & $-0.0346$ & $\phantom{-}0.1399$ & $\phantom{-}0.1114$ \\ 
    \hline
  \end{tabular}
\end{table}

The results of the analysis are shown in Tables \ref{tab:15minute}, \ref{tab:5minute} and \ref{tab:1minute} using three different resolutions of data, that is, 15-minute, 5-minute and 1-minute, respectively.  In particular, for the 5-minute case we have also calculated the Pearson’s coefficients for standard deviations of solar generation ($\rm \sigma_{Solar}$) and wind ($\rm \sigma_{Wind}$), and $\sigma_f$.   
It is interesting to observe that the Pearson’s coefficients for solar are, in general, lower than those for wind and SNSP.  This is to be expected considering the penetration of solar PV at the time of writing (significantly lower than wind).  On the other hand, the Pearson’s coefficients for SNSP is a combination of the solar and wind coefficients.  This is also expected considering the SNSP calculation \cite{10253224}.  However, interestingly when using a 1-minute resolution of data, all the Pearson’s coefficients are reduced and take a negative value.  Using a higher sampling rate, in fact, increases the effect of uncorrelated noise and decreases that of solar variations.  However, if we refer to the Pearson’s coefficients for $\rm \sigma_{Solar}$ and $\rm \sigma_{Wind}$, and $\sigma_f$ in Table \ref{tab:5minute}, then solar generation shows a comparable correlation to wind and coefficients take a positive value.  These results support the conclusion above that solar PV impacts on long-term frequency deviations.  The TSOs are addressing the frequency regulation challenge by reviewing all frequency products as part of SOEF \cite{soef}.

\subsection{Impact on Voltage Magnitude Variations}
\label{sec:voltage}

\begin{figure}[t!]
  \begin{center}
    \resizebox{0.9\linewidth}{!}{\includegraphics{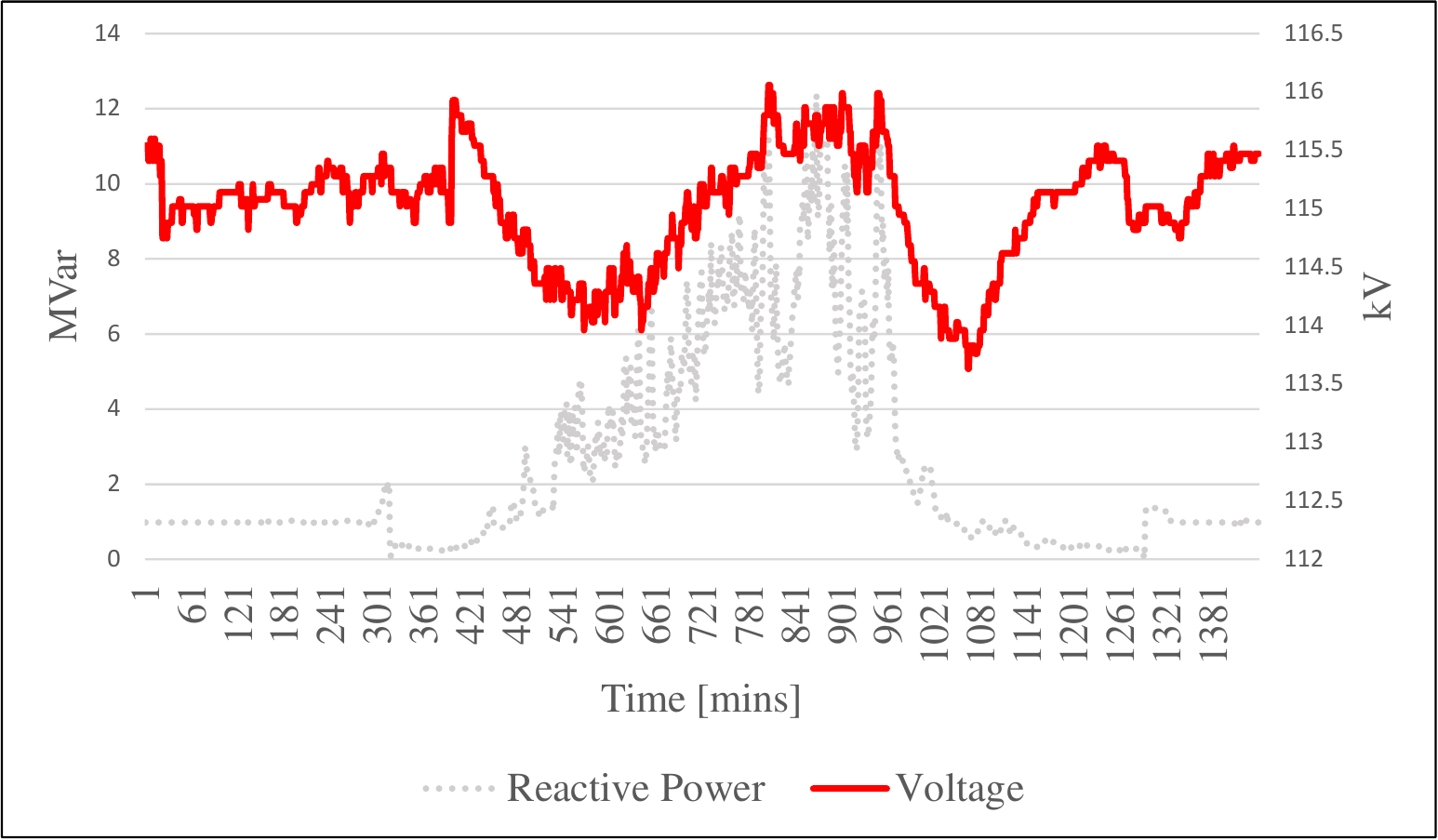}}
    \caption{Daily reactive power and voltage profiles for a relevant solar PV plant.}
    \label{fig:reactive}
  \end{center}
  \vspace*{-0.3cm}
\end{figure}

\begin{figure}[t!]
  \begin{center}
    \resizebox{0.9\linewidth}{!}{\includegraphics{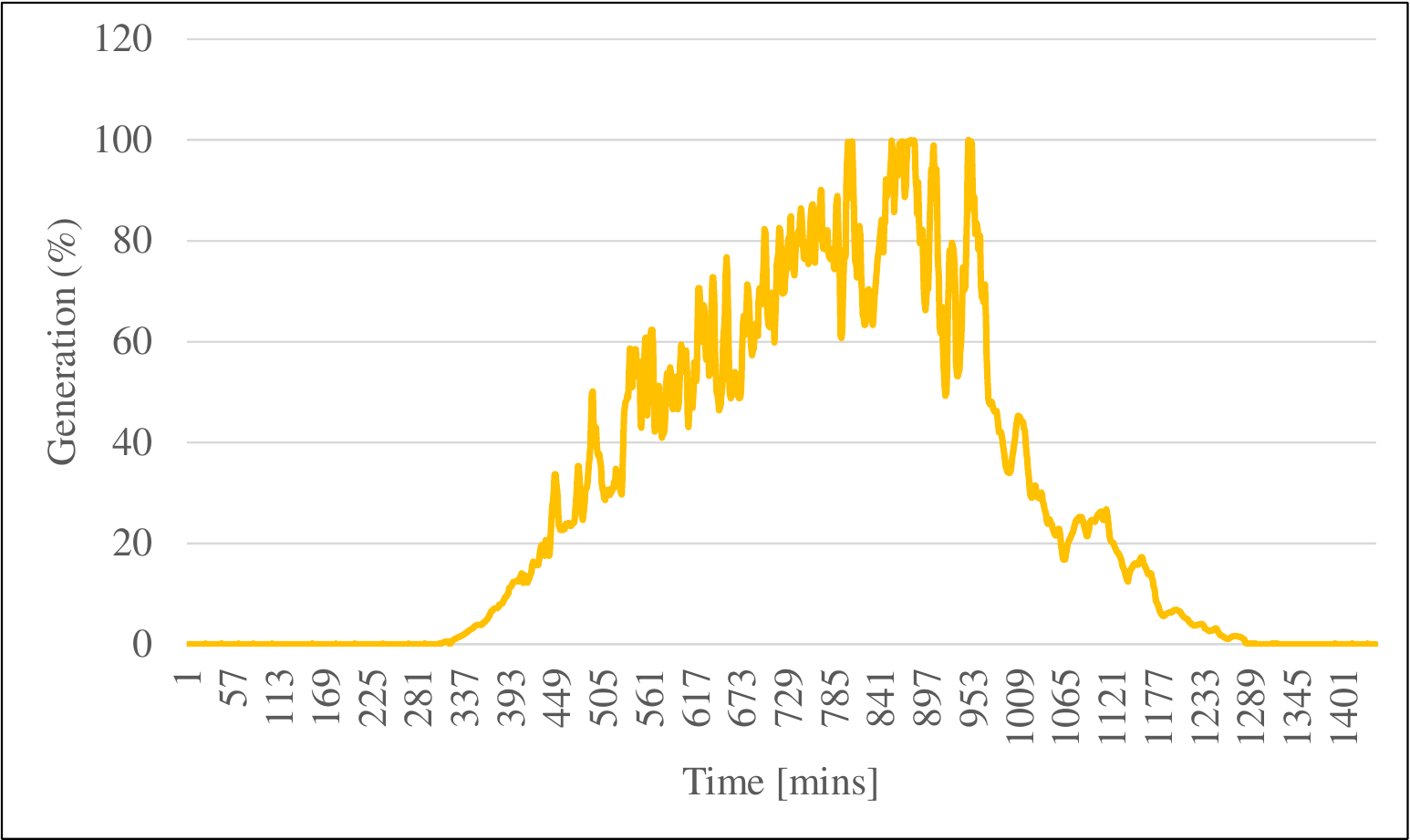}}
    \caption{Daily active power profile for a relevant solar PV plant.}
    \label{fig:active}
  \end{center}
  \vspace*{-0.3cm}
\end{figure}

Common voltage problems caused by high levels of weather-dependent solar PV include voltage fluctuations, unbalance and magnitude variations, respectively \cite{SAMPATHKUMAR2020202}.  In this section, we illustrate the PV-induced voltage magnitude issue (over-voltage) using actual data of a relevant solar PV plant in the AIPS for a particular day.  Figure~\ref{fig:reactive} compares the daily reactive power and voltage magnitude profiles while Fig.~\ref{fig:active} depicts the active power profile of the solar PV plants.  Note that the PV plant is under MVar control mode.  There is a strong correlation between active and reactive power generation and the voltage magnitude profile.  Indeed, without the voltage support capability from the PV plant the situation would have been worse.  Given the expected PV penetration increase in the near future in the AIPS, there is a need for additional voltage support.  Solutions to address these local voltage problems include installing capacitor banks, static VAR compensators and synchronous condensers, among others \cite{SAMPATHKUMAR2020202}.  The TSOs are considering the installation of such voltage-regulating devices in substations to boost the voltage \cite{synchcond}. 

\subsection{Impact on Minimum Operational Demand}
\label{sec:flows}

\begin{figure}[t!]
  \begin{center}
    \resizebox{0.725\linewidth}{!}{\includegraphics{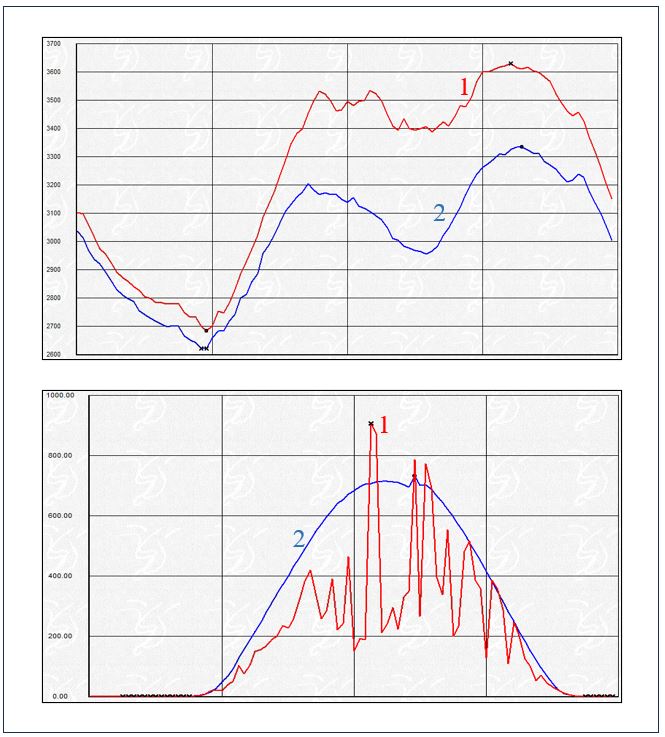}}
    \caption{Impact of solar PV on minimum operational demand.}
    \label{fig:demand}
  \end{center}
  \vspace*{-0.3cm}
\end{figure}

The TSOs rely on the MUON limit to provide a range of system services such as frequency, voltage, and short-circuit contribution and, as such, ensure system security and stability (see Table \ref{tab:constraint}).  However, during days with high distributed PV generation (e.g., rooftop PV), it may be difficult to maintain the MUON limit as the operational demand (as seen from the transmission system) can reduce significantly.  We illustrate this potential issue in the AIPS in Fig.~\ref{fig:demand} using two relevant days and plotting the total demand (upper plot) and the light intensity (lower plot), respectively.   Results show that light intensity and, in turn, high solar PV generation can significantly reduce demand.  A common solution is that of using (long-duration) storage to use the excess PV generation \cite{HUNTER20212077}.  If that is still insufficient, then the TSOs should be able to control and/or disconnect the excess PV generation as a last resort in  emergencies \cite{YILDIZ2023113696}.  In this context, both TSOs and distribution system operators in IE and NI are working towards increased visibility and controllability of these installations to operate the system in a secure and sustainable way \cite{esb}. 

\section{Conclusions}
\label{sec:conclu}

This paper uses operational data to discuss the impact of PV integration on a real-world transmission system, namely the AIPS.  Specifically, the focus is on the impact of solar PV on managing its dispatch down levels, frequency, and voltage variations, as well as maintaining a minimum operational demand.  The case study shows that the current PV penetration in the AIPS, despite still being low compared to wind power, is challenging the operation of the AIPS.  In particular, keeping frequency and voltage within operational limits is becoming an emerging challenge.  The TSOs are working together towards addressing all the above challenges.



\end{document}